\documentclass[aps,prl,twocolumn,showkeys,showpacs,amsmath,amssymb,nofootinbib,superscriptaddress,floatfix,reprint,longbibliography]{revtex4-1}
\usepackage[dvips]{graphicx}
\usepackage{latexsym}
\usepackage{amsmath}
\usepackage{amsfonts}
\usepackage{amssymb}
\usepackage{color}
\usepackage{txfonts}
\usepackage{float}
\usepackage{url}
\usepackage[colorlinks=true, urlcolor=blue, linkcolor=blue, citecolor=blue, pdftex]{hyperref}
\usepackage{ulem}
\usepackage{graphicx}
\usepackage{extpfeil}
\usepackage{subfigure}
\usepackage{physics}
\DeclareUnicodeCharacter{2212}{-}

\begin{document}
	\newcommand{\fig}[2]{\includegraphics[width=#1]{#2}}
	\newcommand{\la}{{\langle}}
	\newcommand{\ra}{{\rangle}}
	\newcommand{\dg}{{\dagger}}
	\newcommand{\upa}{{\uparrow}}
	\newcommand{\dna}{{\downarrow}}
	\newcommand{\ab}{{\alpha\beta}}
	\newcommand{\ias}{{i\alpha\sigma}}
	\newcommand{\ibs}{{i\beta\sigma}}
	\newcommand{\hH}{\hat{H}}
	\newcommand{\hn}{\hat{n}}
	\newcommand{\hc}{{\hat{\chi}}}
	\newcommand{\hU}{{\hat{U}}}
	\newcommand{\hV}{{\hat{V}}}
	\newcommand{\br}{{\bf r}}
	\newcommand{\bk}{{{\bf k}}}
        \newcommand{\bbk}{{\textit{\textbf{k}}}}
        \newcommand{\bbK}{{\textit{\textbf{K}}}}
	\newcommand{\bq}{{{\bf q}}}
	\def\gsim{~\rlap{$>$}{\lower 1.0ex\hbox{$\sim$}}}
	\setlength{\unitlength}{1mm}
	\newcommand{\pprl}{Phys. Rev. Lett. \ }
	\newcommand{\pprb}{Phys. Rev. {B}}
         \newcommand{\dx}{$d_{x^2-y^2}$}
         \newcommand{\dz}{$d_{z^2}$ }
        \newcommand{\LNO}{La$_3$Ni$_2$O$_7$ }

\title {The electronic structure and disorder effect of La$_3$Ni$_2$O$_{7}$ superconductor}

\author{Yuxin Wang}
\affiliation{Beijing National Laboratory for Condensed Matter Physics and Institute of Physics,
	Chinese Academy of Sciences, Beijing 100190, China}
\affiliation{School of Physical Sciences, University of Chinese Academy of Sciences, Beijing 100190, China}

\author{Yi Zhang}
\email{zhangyi821@shu.edu.cn}
\affiliation{Department of Physics and Institute for Quantum Science and Technology, Shanghai University, Shanghai 200444, China}
\affiliation{Shanghai Key Laboratory of High Temperature Superconductors and International Center of Quantum and Molecular Structures, Shanghai University, Shanghai 200444, China}     

\author{Kun Jiang}
\email{jiangkun@iphy.ac.cn}
\affiliation{Beijing National Laboratory for Condensed Matter Physics and Institute of Physics,
	Chinese Academy of Sciences, Beijing 100190, China}
\affiliation{School of Physical Sciences, University of Chinese Academy of Sciences, Beijing 100190, China}

\date{\today}

\begin{abstract}
Determining the electronic structure of La$_3$Ni$_2$O$_7$ is an essential step towards uncovering their superconducting mechanism.
It is widely believed that the bilayer apical oxygens play an important role in the bilayer La$_3$Ni$_2$O$_7$ electronic structure.
Applying the hybrid exchange-correlation functionals, we obtain a more accurate electronic structure of La$_3$Ni$_2$O$_7$ at its high-pressure phase, where the bonding \dz band is below the Fermi level owing to apical oxygen. The symmetry properties of this electronic structure and its corresponding tight-binding model are further analyzed. We find the antisymmetric part is highly entangled leading to a minimal nearly degenerate two-orbital model. Then, the apical oxygen vacancies effect is studied using the dynamical cluster approximation. 
This disorder effect strongly destroys the antisymmetric $\beta$ Fermi surface leading to the possible disappearance of superconductivity. 
\end{abstract}



\maketitle

The recently discovered nickelate superconductor, La$_3$Ni$_2$O$_7$, under high-pressure conditions, significantly broadens our understanding of high-temperature superconductivity \cite{meng_wang,hwang,chen_zhuoyu}. However, the intricate bilayer structure of La$_3$Ni$_2$O$_7$, the presence of multiple competing phases, and the requirement of high-pressure environments pose substantial challenges for both experimental and theoretical investigations \cite{chengjg_crystal,hwang,chen_zhuoyu,meng_wang,chengjg,yuanhq,chengjg_crystal,chengjg_poly,yaodx,dagotto1,wangqh,Kuroki,guyh,zhanggm,werner,yangf,wucj,sugang,dagotto2,yangyf2,ryee2023critical,kun_cpl,taoxiang,hepting_PhysRevLett.133.146002,Mitchell_mixing,ueki2024phase}.
In particular, understanding the electronic structure of \LNO is crucial to unraveling its superconducting properties. Exploring how the electronic structure evolves under varying conditions is a critical step toward establishing a minimal model and uncovering the mechanism behind its superconductivity.
For example, it is widely known that the apical oxygen between two bilayer Ni atoms plays a special role in their properties \cite{meng_wang,chengjg_crystal,oxygen_vacancies}. 
Therefore, understanding the role of apical oxygen and its associated vacancies are important for further experimental exploration.

In this work, we focus on the \LNO electronic structure at the superconducting phase. The latest experimental results indicate that when the pressure is approximately 18 GPa, the crystal structure adopts the $Fmmm$ space group, at which point the superconducting transition temperature reaches its maximum. When the pressure exceeds 40 GPa, the space group of the crystal structure transitions to $I4/mmm$ \cite{li2024pressure}.
Therefore, we can start from the electronic structure of \LNO at its $Fmmm$ structure.  We apply the density functional theory (DFT) calculation with
hybrid functional, which more accurately treats the Coulomb potential.
This method has given a faithful description for the ambient pressure \LNO electronic and magnetic structures \cite{yxwang_PhysRevB.110.205122,rixs,zhouxj}.
The result is consistent with our previous findings, where the $d_{z^2}$ bonding band is always below the Fermi level. Then, the influence of apical oxygen and the random disorder effect of apical oxygen is explored by the dynamical cluster approximation (DCA)~\cite{dca1,dca2,dca_review}.

\begin{figure}
	\begin{center}
		\fig{3.4in}{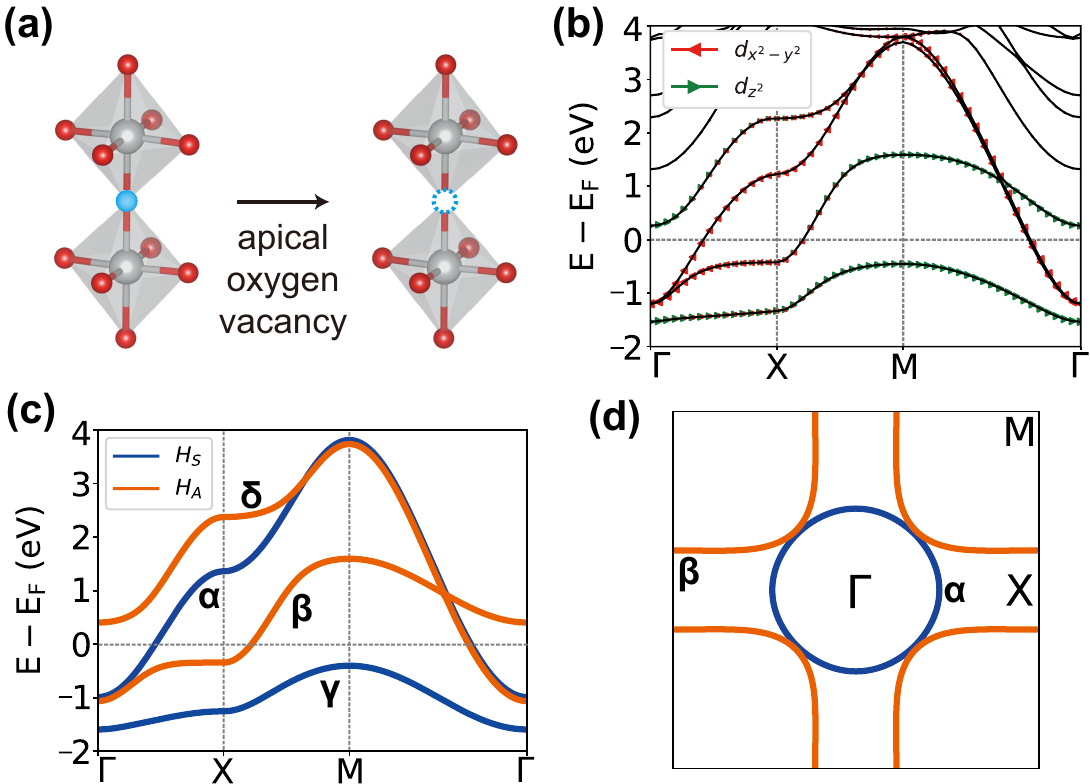}
		\caption{(a) The crystal structure of \LNO at its high-pressure (HP) phase. The oxygen vacancies are located at the positions of apical oxygen. (b) The corresponding orbital-resolved DFT band structure. The red dots and the green dots denote the $d_{x^2-y^2}$ orbital and $d_{z^2}$ orbital, respectively. (c) The tight-binding bands are divided into $H_{S}$ sector (blue lines) and $H_{A}$ sector (orange lines). We label the four bands as $\alpha$, $\beta$, $\gamma$ and $\delta$. (d) The Fermi surfaces of the corresponding TB model with an antisymmetric $\beta$-FS and a symmetric $\alpha$-FS.
			\label{fig1}}
	\end{center}
\end{figure}

We apply the standard first-principles density functional theory calculations on \LNO for determining its electronic structures. Most of the aforementioned DFT calculations employ the generalized gradient approximation (GGA) for the exchange-correlation functional, often using its Perdew-Burke-Ernzerhof (PBE) variant \cite{perdew1996generalized}. However, it is well established that the choice of exchange-correlation functional $E_{xc}$ plays a critical role in accurately predicting properties such as band gaps and binding energies \cite{martin2020electronic}. For instance, GGA functionals are known to systematically underestimate the band gaps of semiconductors \cite{martin2020electronic,exchange_correlation}. In this work, we adopt a more accurate hybrid exchange-correlation functional to address these limitations. 

Our DFT calculations are performed using the Vienna Ab-initio Simulation Package (VASP) \cite{kresse1996efficient} with the projector augmented wave (PAW) method \cite{kresse1999ultrasoft}. To achieve greater accuracy, we employ hybrid functionals that combine Hartree-Fock (HF) and Kohn-Sham (KS) theories \cite{becke1993density}, which typically outperform the semilocal GGA method \cite{exchange_correlation,Jacob_ladder}. Specifically, we utilize the widely adopted HSE06 hybrid functional \cite{krukau2006influence}. The cutoff energy for the plane-wave basis set is set at 500 eV, and the energy convergence criterion is $10^{-6}$ eV. To optimize computational efficiency, all calculations are conducted using the primitive cell. A $\Gamma$-centered $7\times7\times7$ k-point mesh is used for sampling the Brillouin zone. 

\begin{table}
	\begin{tabular}{ccccc}
		\hline \hline 
		
		$t_{1}^{x}$ & $t_{1}^{z}$ & $t_{2}^{x}$ & $t_{2}^{z}$ & $t_{3}^{xz}$  \tabularnewline\hline 
		-0.6003 & -0.149 & 0.0391 & -0.0007 & 0.2679 \tabularnewline \hline 
		
		$t_{\bot}^{x}$ & $t_{\bot}^{z}$ & $t_{4}^{xz}$ & $\epsilon^{x}$ & $\epsilon^{z}$  \tabularnewline \hline 
		0.038 & -0.999 & -0.072 &  1.2193 & 0.0048 \tabularnewline
		
		\hline \hline
	\end{tabular}
    
    \vspace{0.5cm}
    
    \begin{tabular}{cccccccc}
		\hline \hline 
	sector	 &$t_{1}^{x}$ & $t_{1}^{z}$ & $t_{2}^{x}$ & $t_{2}^{z}$ & $t_{3}^{xz}$ & $\epsilon^{x}$ & $\epsilon^{z}$   \tabularnewline\hline 
	$H_{S}$&	-0.6003 & -0.149 & 0.0391 & -0.0007 & 0.1959  &  1.2573  & -0.9942 \tabularnewline \hline
	$H_{A}$&	-0.6003 & -0.149 & 0.0391 & -0.0007 & 0.3399  &  1.1813 &  1.0038  \tabularnewline 
		\hline \hline
	\end{tabular}
    
	\caption{\label{tab:hop1}The hopping parameters of the TB Hamiltonian  $H(\textbf{k})$ in the HP phase in unit of eV.  $\epsilon^{x},\epsilon^z$ are site energies for Ni $d_{x}$ and $d_{z}$ orbitals, respectively. The top one corresponds to the Eq. \ref{eq:tb-lp}, and the bottom one corresponds to the Eq. \ref{eq:tb-lp-sa}.  
	}
 \label{HP_TB}
\end{table}

The orbital-resolved band structures for the HP phase without oxygen vacancy disorder, calculated using the HSE06 functional, are shown in Fig. \ref{fig1}(b). Given that the material only exhibits $C_2$ symmetry, the $x$- and $y$-directions are inequivalent. However, the disparity between them is negligible. We can reasonably approximate the $x$- and $y$-directions as equivalent, treating the lattice as effectively square. Compared to the GGA-PBE results \cite{yaodx,guyh}, the bonding $d_{z^2}$ bands are pushed significantly below  $E_F$ by about $452$ meV ($M$ point in Fig. \ref{fig1}(b)). This enhancement is owing to enlarged crystal field splitting and perpendicular hopping mediated by the apical oxygens.
Recently, the superconducting \LNO thin films have been successfully synthesized \cite{hwang,chen_zhuoyu}. It is interestingly that the in-plane lattice constants $a,b$ are suppressed to $3.71 \AA$ instead of $3.83 \AA$ in high-pressure. The Jahn-Teller distortion between $d_{x^2-y^2}$ and \dz is largely enhanced. This effect also leads to bonding \dz band below the Fermi level even using the GGA-PBE exchange-correlation functional, as discussed in the Appendix (See Fig. \ref{figS1}). Therefore, both the high-pressure phase and thin film phase show a similar band structure, and this is another strong reason supporting the correctness of the hybrid functional.

Focusing on the partially occupied $e_g$ orbitals from the DFT calculations, we use the Wannier90 code \cite{mostofi2008wannier90,marzari2012maximally} to fit the $e_g$ bands and extract the tight-binding (TB) model parameters. Given that there are two Ni atoms per unit cell, the resulting TB model is a 4-band system. Its Hamiltonian, $H_{HP}$, expressed in the basis  $(d_{t\textbf{k}}^{x},d_{t\textbf{k}}^{z},d_{b\textbf{k}}^{x},d_{b\textbf{k}}^{z})$ (the spin index is omitted here), is written as:
\begin{align}
H_{HP} & ({\textbf{k}})=\left(\begin{array}{cc}
H_{t}({\textbf{k}}) & H_{\perp}({\textbf{k}})\\
H_{\perp}^{\dagger}({\textbf{k}}) & H_{b}({\textbf{k}})
\end{array}\right),\label{eq:tb-lp}
\end{align}
where $H_{b}({\textbf{k}})=H_{t}({\textbf{k}})$ and $H_{\perp}({\textbf{k}})=H_{\perp}^{\dagger}({\textbf{k}})$, and they are both $2\times 2$ block matrices.
They take the structures of
\begin{align}
H_{t}({\textbf{k}})=\left(\begin{array}{cccc}
T_{{\textbf{k}}}^{x} & V_{{\textbf{k}}}\\
V_{{\textbf{k}}} & T_{{\textbf{k}}}^{z}
\end{array}\right), \label{ht}
\end{align}
and
\begin{align}
H_{\perp}({\textbf{k}})=\left(\begin{array}{cc}
t_{\bot}^{x} & V_{{\textbf{k}}}^{\prime}\\
V_{{\textbf{k}}}^{\prime} & t_{\bot}^{z}
\end{array}\right). 
\end{align}
Here,
$T_{{\textbf{k}}}^{x/z}=t_{1}^{x/z}\gamma_k+t_{2}^{x/z}\alpha_k+\epsilon^{x/z}$, $V_{\textbf{k}}=t_{3}^{xz}\beta_k$, $V_{\textbf{k}}^{\prime}=t_{4}^{xz}\beta_k$ with $\gamma_k=2(\cos k_x+\cos k_y)$, $\alpha_k=4\cos k_x\cos k_y$, $\beta_k=2(\cos k_x-\cos k_y)$. The corresponding on-site energies and hopping parameters are given in Table. \ref{HP_TB}. The dispersion of the TB $e_g$ bands is compared to the DFT calculations in the Appendix (See Fig. \ref{figS2}(a)), which confirms that the TB model provides a faithful description of the low-energy electron structure.

\begin{figure}
	\begin{center}
		\fig{3.4in}{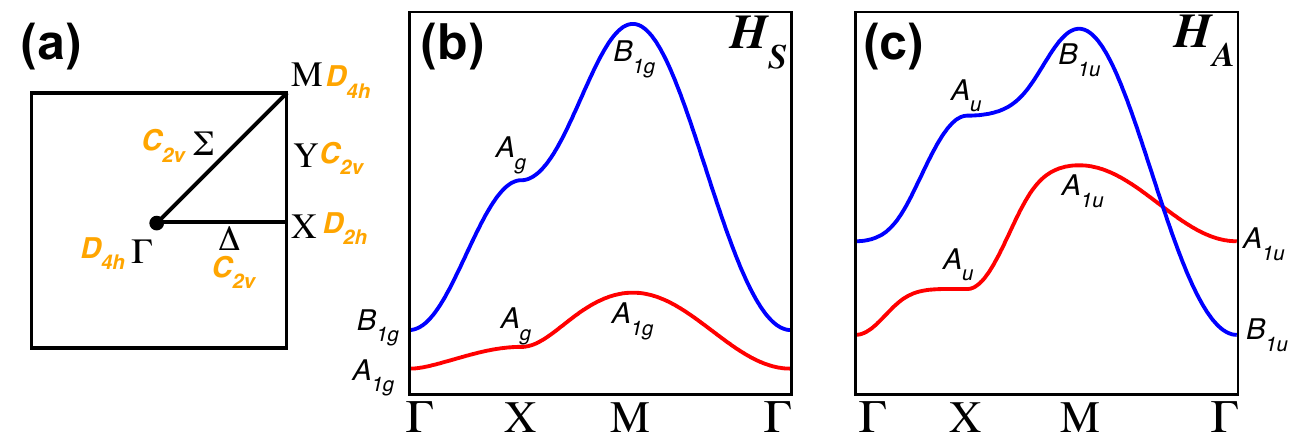}
		\caption{(a) The point group of the wavevector $k$ along the high-symmetry lines and high-symmetry points.
       We adopt the approximate $D_{4h}$ point group. (b) The irreducible representation of $H_{S}$ sector at each high-symmetry point.
       The compatibility of different irreducible representations is satisfied. (c) The irreducible representation of $H_{A}$ sector at each high-symmetry point. Especially, the $A_{1u}$ and $B_{1u}$ representations switch their positions at $\Gamma$ and $M$ points. 
			\label{fig2}}
	\end{center}
\end{figure}

From a symmetry point of view, we can use the inversion eigenbasis centered at the shared apical oxygen \cite{kun_cpl}
\begin{eqnarray}
	\psi_{S}^{\eta}&=&(d_{t}^{\eta}+d_{b}^{\eta})/\sqrt{2} \\
	\psi_{A}^{\eta}&=&(d_{t}^{\eta}-d_{b}^{\eta})/\sqrt{2}.
\end{eqnarray}
They correspond to the symmetric and antisymmetric molecular orbitals that block diagonalize $H_{HP} ({\textbf{k}})$ into 
\begin{eqnarray}
		H_{HP}(\textbf{k})=\left(\begin{array}{cc} 
		H_S(\textbf{k}) & 0 \\
		0 & H_A(\textbf{k})
	\end{array}\right), \label{eq:tb-lp-sa}
\end{eqnarray}
where $H_{S/A}(\textbf{k})=H_{t}(\textbf{k})\pm H_{\perp}(\textbf{k})$, whose matrix elements can take the same form as $H_{t}(\textbf{k})$ in Eq. \ref{ht}. That is,
\begin{align}
H_{S/A}({\textbf{k}})=\left(\begin{array}{cccc}
T_{{\textbf{k}},S/A}^{x} & V_{{\textbf{k}},S/A}\\
V_{{\textbf{k}},S/A} & T_{{\textbf{k}},S/A}^{z}
\end{array}\right), 
\end{align}
with $T_{{\textbf{k}},S/A}^{x/z}=t_{1,S/A}^{x/z}\gamma_k+t_{2,S/A}^{x/z}\alpha_k+\epsilon^{x/z}_{S/A}$ and $V_{\textbf{k},S/A}=t_{3,S/A}^{xz}\beta_k$. The corresponding parameters are also given in Table. \ref{HP_TB}.
The TB electronic structure divided into two $\psi_{S}$ bands (blue lines) and two $\psi_{A}$ bands (orange lines) is plotted in Fig. \ref{fig1}(c). There are two bands crossing $E_F$, labeled by $\alpha$, and $\beta$, separated from an unoccupied $\delta$ band and a fully occupied $\gamma$ band. The Fermi surfaces (FSs) consist of one electron pocket ($\alpha$ band) around the $\Gamma$ point and one hole pocket ($\beta$ band) around the $M$ points in the Brillouin zone as shown in Fig. \ref{fig1}(d).

Before a detailed discussion, it is useful to look at the symmetry properties of each band. As discussed above, the space group of \LNO is $Fmmm$, similar to Bi-2212 cuprates \cite{2212-PhysRevB.38.893,2212-PhysRevLett.60.1174}. Although the point group is $D_{2h}$, it is safe to use $D_{4h}$ to analyze the band symmetry as in cuprates. Hence, the point group of the wavevector $k$ along the high-symmetry lines and high-symmetry points are listed in Fig. \ref{fig2}(a). The group of wavevector is $D_{4h}$ for the $\Gamma$ and $M$ points while it becomes $D_{2h}$ for $X$ point.

Then, the group representations for each band and their compatibility can be calculated. Looking at the symmetric sector $H_S$, the two bands are separated as plotted in Fig. \ref{fig2}(b). The lower band with $A_{1g}$ at $\Gamma$ is mainly formed by $d_{z^2}$ while the upper band with $B_{1g}$ is mainly from $d_{x^2-y^2}$. On the other hand, the symmetry properties of $H_{A}$ are more complicated as shown in Fig. \ref{fig2}(c).
Although the $A_{1u}$ band is slightly higher than the $B_{1u}$ band at $\Gamma$, they switch their positions at $M$ point.
This feature can be easily tested along the $\Gamma$-$M$ direction, where the overlap matrix between two orbitals vanishes. The complicated representation connection makes the $d_{x^2-y^2}$ and \dz orbital largely entangled. 
Therefore, although $H_A$ only contributes to a single $\beta$-FS, it is difficult to use a single Wannier orbital for this FS. 


We notice that the $\epsilon_x$ and $\epsilon_z$ are almost degenerate for $H_A$. This fact is owing to the cancelation between Jahn-teller distortion and interlayer hopping. It is obvious that the $\beta$-FS shows a similar contour as the cuprate superconductor. 
As we point out at Refs. \cite{kun_cpl,wang2024selfdopedmolecularmottinsulator}, the essential pairing comes from this antisymmetric FS. 
Therefore, the minimal effective model for $H_A$ is a nearly degenerate two-orbital model \cite{wang2024selfdopedmolecularmottinsulator}. The self-doping $H_A$ leads to the high-temperature superconductivity in \LNO \cite{wang2024selfdopedmolecularmottinsulator}.

As discussed above, the apical oxygen vacancies are believed to impact the superconducting phase of the system greatly.
Hence, the target system becomes La$_3$Ni$_2$O$_{7-\delta}$ with $\delta$ the concentration of the apical oxygen vacancies per unit cell.
The oxygen vacancies play two significant roles. First, as the apical oxygen has a chemical valence of O$^{2-}$ and contributes two holes to the unit cell, its vacancies, with a concentration $\delta$, result in an electron doping level of $x = 2\delta$. Second, the random distribution of these vacancies introduces disorder into the system.
We use DCA~\cite{dca1,dca2,dca_review} to study the disorder effect induced by the random vacancies, which maps the disordered lattice to a finite cluster embedded in an effective medium described by a non-local self-energy which is then determined self-consistently.
Since the apical oxygen is located between the two Ni layers, its vacancy primarily reduces the coupling strength between these layers.
Consequently, the disordered system with randomly distributed oxygen vacancies can be modeled using a four-band Anderson model with random intracell potentials. The Hamiltonian can be written as
\begin{equation}
    H=\sum_{i,j}\sum_{\alpha,\beta=1}^{4} (h_{ij}^{\alpha\beta}-\mu\delta_{\alpha\beta})c_{i\alpha}^{\dagger}c_{i\beta} +\sum_{i}\sum_{\alpha,\beta=1}^{4}V_i^{\alpha\beta}c_{i\alpha}^{\dagger}c_{i\beta} \ .
\end{equation}
The first term describes the band of La$_3$Ni$_2$O$_7$ with $h_{ij}^{\alpha\beta}$ the real space version of the lattice model $H_{HP}({\textbf{k}})$ defined in Eq. ~\ref{eq:tb-lp-sa} and $\mu$ the chemical potential.
The labels $i$, $j$ are the indices for the unit cell and $\alpha$, $\beta$ are the combined indices for the layers and the two $e_g$ orbitals so that $c_{i\alpha}^{\dagger}$ ($c_{i\alpha}$) correspond to the creation (annihilation) operators for the orbitals $(d_{S}^{x},d_{S}^{z},d_{A}^{x},d_{A}^{z})$ on the cell $i$.
The second term represents the disorder, modeled by the intracell potential $V_i^{\alpha\beta}$, which is randomly distributed according to an independent binary probability distribution function of the form~\cite{mbdca}
\begin{equation}
    P(V_i^{\alpha\beta})=\delta * \delta(V_i^{\alpha\beta}-V_{vac}) +(1-\delta) * \delta(V_i^{\alpha\beta})
    \label{eq:imp}
\end{equation}
Here we use the impurity potential $V_{vac}$ to remove the contribution of intracell interlayer coupling $t_{\perp}^x$ and $t_{\perp}^z$ to $\epsilon^{x}$ and $\epsilon^z$ in the eigenbasis described above, which should capture the main effect of the oxygen vacancies.

\begin{figure}
	\begin{center}
		\fig{3.4in}{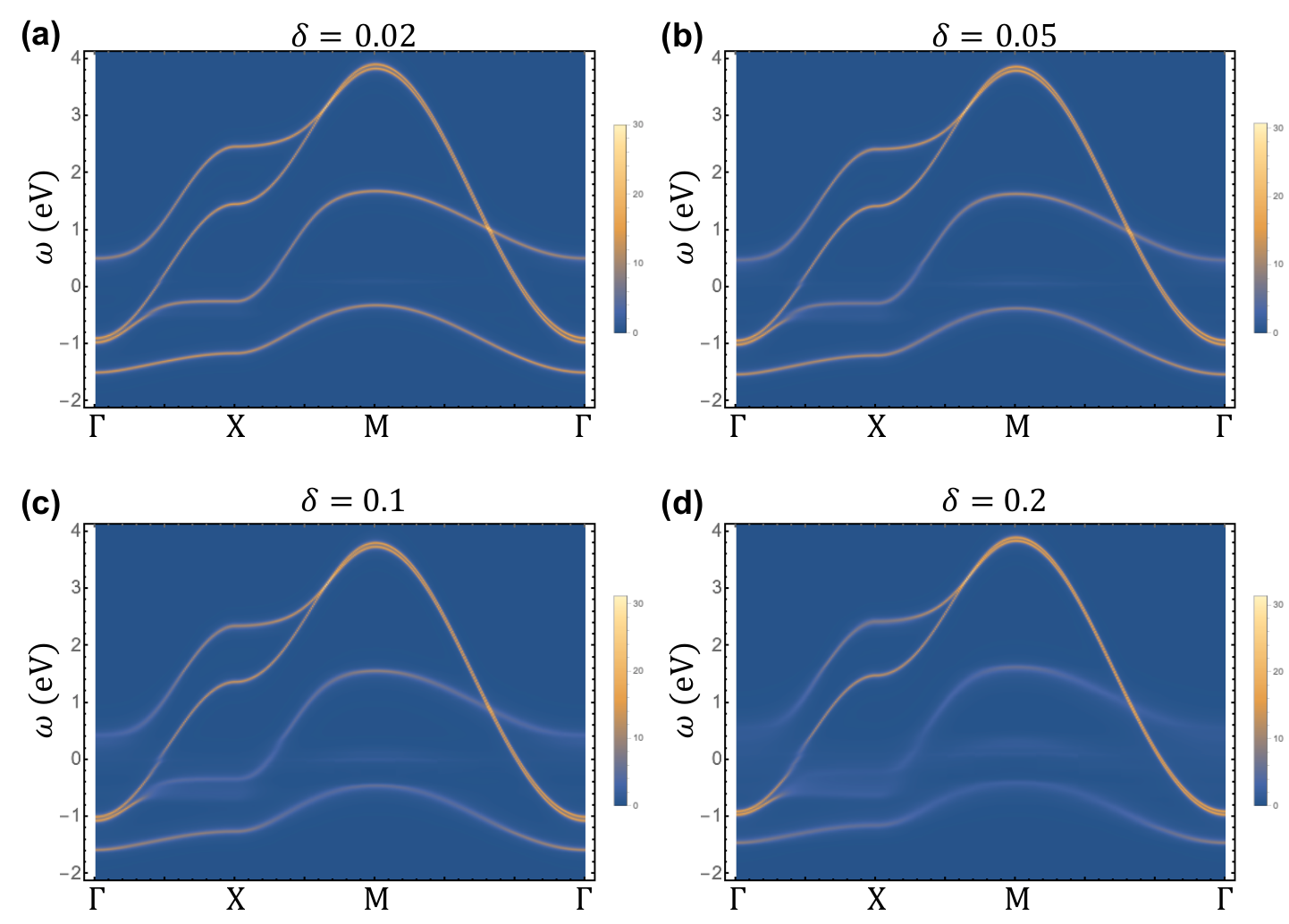}
		\caption{The calculated spectrum function $A(\bbk,\omega)$ using the lattice Green's function with the DCA self-energy $\Sigma(\bbK,\omega)$ along the high-symmetry lines and high-symmetry points for various concentrations of the apical oxygen vacancies $\delta$. Here $\omega$ is measured from the chemical potential. 
        The impurity potential is modeled by assuming the apical oxygen vacancy completely removes the interlayer intracell coupling $t_{\perp}^{x}$ and $t_{\perp}^{z}$  and the calculation is performed at zero temperature.
	\label{fig3}}
	\end{center}
\end{figure}

In the calculation of DCA, the original lattice model is mapped onto a cluster of size $N_c$ with periodic boundary conditions, embedded in an effective medium. Consequently, the first Brillouin zone is divided into $N_c$ coarse-grained cells~\cite{dca_review}, each centered around a momentum labeled by $\bbK$ and surrounded by $k$ points labeled as $\tilde{\bbk}$ within the cell. Thus, all $k$ points can be expressed as $\bbk = \bbK + \tilde{\bbk}$.
The detailed formulation of the multi-band DCA can be found in Ref.~\onlinecite{mbdca}.
In the cluster solver of the DCA, we perform stochastic sampling of configurations based on the binary distribution of the impurity potential described in Eq.~\ref{eq:imp}. For all calculations, we use a cluster size of $N_c = 100$ and average over 2560 disorder configurations, which is sufficient to produce reliable results.
After the DCA calculation is finished, we use the obtained cluster self-energy $\Sigma(K,\omega)$ to construct the lattice Green's function $G_{lat}(k,\omega)$ as
\begin{equation}
    G_{lat}(\bbk,\omega)=[(\omega+\mu)\mathbb{I}-H_{HP}(\bbk)-\Sigma(\bbK,\omega)]^{-1}
\end{equation}
where $\mathbb{I}$ is the identity matrix in orbital basis and $\bbk = \bbK + \tilde{\bbk}$, the chemical potential $\mu$ is determined according to the doping level $x=2\delta$ introduced by the oxygen vacancies and $\omega$ is measured from the chemical potential.
We next calculate the spectrum function $A(k,\omega)$ as
\begin{equation}
    A(\bbk,\omega)=-\frac{1}{\pi}\text{Im}[\Tr G_{lat}(\bbk,\omega)]
\end{equation}
which reflects the disorder effect due to the random oxygen vacancies. The DCA calculation is performed at zero temperature.

\begin{figure}
	\begin{center}
		\fig{3.4in}{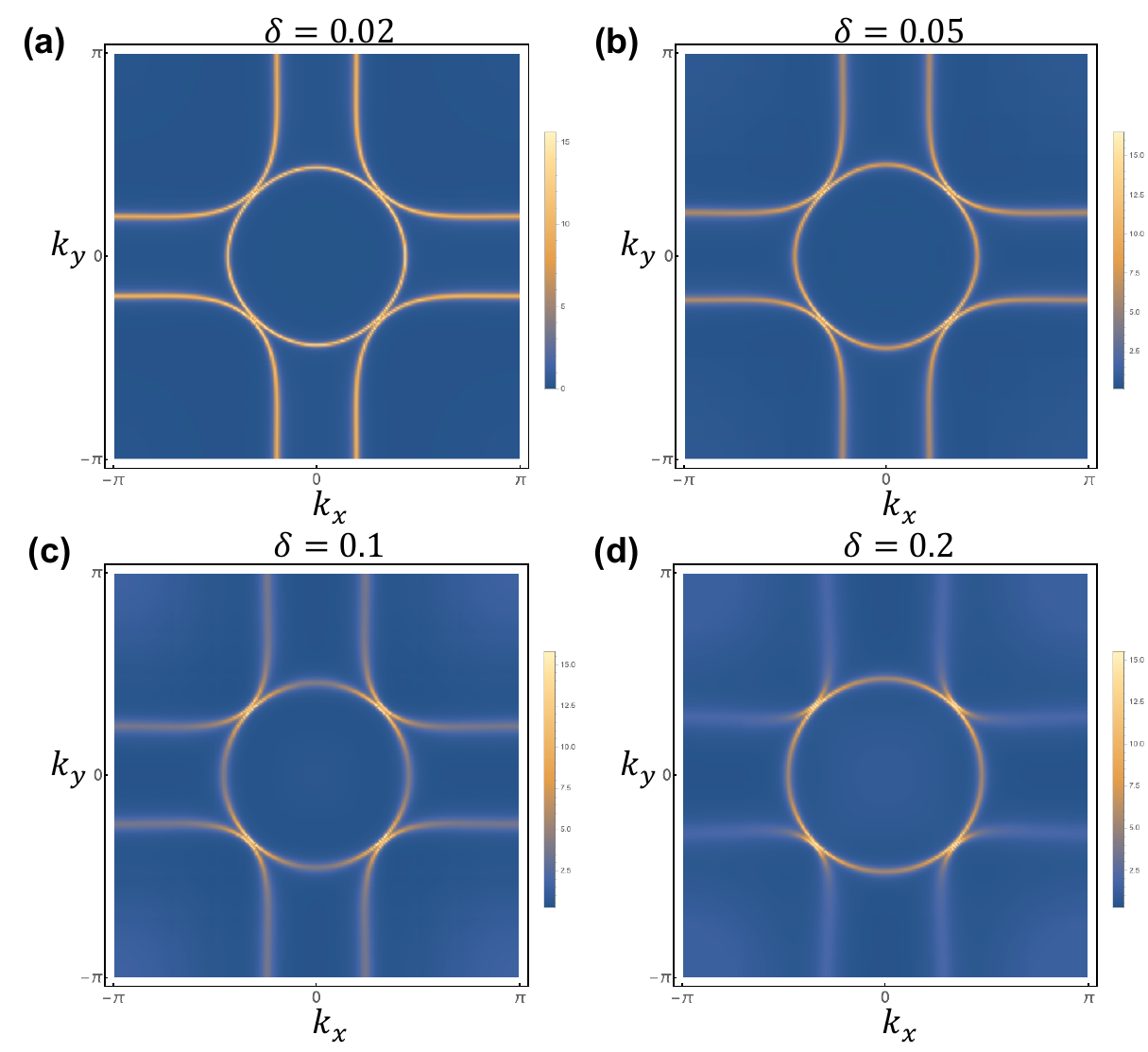}
		\caption{The calculated spectrum function at the chemical potential $A(\bbk,0)$ using the lattice Green's function with the DCA self-energy $\Sigma(\bbK,0)$ in the whole Brillouin zone for various concentrations of the apical oxygen vacancies $\delta$.
        As the vacancies increase, the $\beta$-FS becomes more vague while the $\alpha$-FS remains sharp.
        The impurity potential is modeled by assuming the apical oxygen vacancy completely removes the interlayer intracell coupling $t_{\perp}^{x}$ and $t_{\perp}^{z}$  and the calculation is performed at zero temperature.
			\label{fig4}}
	\end{center}
\end{figure}

The calculated spectrum functions $A(\bbk,\omega)$ for various concentrations of the oxygen vacancies along the momentum path defined in Figs.~\ref{fig1} and \ref{fig2} are shown in Fig.~\ref{fig3}.
We can see that for the two bands close to the Fermi level ($\alpha$ and $\beta$ band), the spectrum function of the $\beta$ band decreases much faster than that of the $\alpha$ band as the concentration of the oxygen vacancies increases, indicating that the $\beta$ band is more fragile against the disorder introduced by the oxygen vacancies.
This result is expected, as the primary effect of oxygen vacancies is to disrupt the coupling between the two Ni layers. However, the interlayer coupling for the $d_{z^2}$  orbital is much stronger than that for the $d_{x^2-y^2}$  orbital, as shown in Table~\ref{HP_TB}. Given that the  $\beta$ band has a significant contribution from the  $d_{z^2}$ orbital, while the  $\alpha$ band is primarily composed of the  $d_{x^2-y^2}$  orbital, it follows that the  $\beta$ band is more sensitive to disorder induced by the oxygen vacancies. This is also reflected in the spectrum function at the Fermi level, which is shown in Fig.~\ref{fig4}. We can see that the spectrum function on the Fermi surface of the $\beta$ band is destroyed quickly as the oxygen vacancy increases while that for the Fermi surface of the $\alpha$ band is relatively robust against the disorder.
As we state above and in Ref. \cite{wang2024selfdopedmolecularmottinsulator}, the superconductivity of \LNO is driven by the $\beta$-FS and antisymmetric bands. Therefore, the existence of $\beta$ under disorder, especially the apical oxygen vacancies, becomes critical for superconductivity. Based on DCA calculations, the apical oxygen vacancies strongly ruin the $\beta$-FS resulting in destroying the superconductivity.

Although direct ARPES measurements on high-pressure phases are currently unavailable, two indirect approaches can still be employed to verify the validity of our proposed scenario. First, for \LNO thin films grown on SrLaAlO$_{4}$ (SLAO) substrates—which exhibit superconductivity under ambient pressure \cite{hwang,chen_zhuoyu}—we can directly measure the evolution of the $\beta$-Fermi surface using ARPES when oxygen vacancies are introduced. Second, the superconducting pairing symmetry can be experimentally determined. Prior studies have debated this extensively, proposing both $s^{\pm}$-wave and $d$-wave possibilities \cite{yaodx,wangqh,guyh,yangf,wucj,dagotto2,ryee2023critical,kun_cpl}. If experiments confirm a $d$-wave symmetry, this would strongly suggest that the $\beta$-band dominates the superconducting pairing, thereby supporting the conclusions of our work.

In summary, we carry out an electronic structure calculation on high-pressure \LNO. Using the hybrid exchange-correlation functional, the crystal field and interlayer coupling through apical oxygen are improved leading to the bonding $d_{z^2}$ below the Fermi level.
We construct the bilayer two-orbital tight-binding model using wannierization. This model is further classified into symmetric $H_{S}$ and antisymmetric $H_{A}$ sectors. Using the symmetry analysis, we find the two $H_{A}$ bands are strongly entangled. The minimal model for $H_{A}$ is a nearly degenerate two-orbital model. 
Then, the apical oxygen vacancy disorder effect is investigated using dynamical cluster approximation. The oxygen vacancies have strong effects on the $\beta$-FS. Therefore, increasing the vacancy disorder will destroy the $\beta$-FS quickly and subsequently kill the superconductivity. We hope our findings could provide a new insight into \LNO electronic structure and the corresponding disorder effect. 

\textit{Acknowledgement:}
We acknowledge the support by the National Natural Science Foundation of China (Grant NSFC-12494590,  No. NSFC-12174428, and No. NSFC-12274279), the New Cornerstone Investigator Program, and the Chinese Academy of Sciences Project for Young Scientists in Basic Research (2022YSBR-048). 

\bibliographystyle{iopart-num.bst}
\bibliography{ref}

\clearpage
\appendix
\begin{center}
\textbf{\large Appendix}
\end{center}
\renewcommand{\thetable}{S\arabic{table}}
\setcounter{table}{0}
\renewcommand{\thefigure}{S\arabic{figure}}
\setcounter{figure}{0}

\section{A. The dependence of the GGA-PBE electronic structure on the in-plane lattice constants}

\begin{figure}
	\begin{center}
		\fig{3.4in}{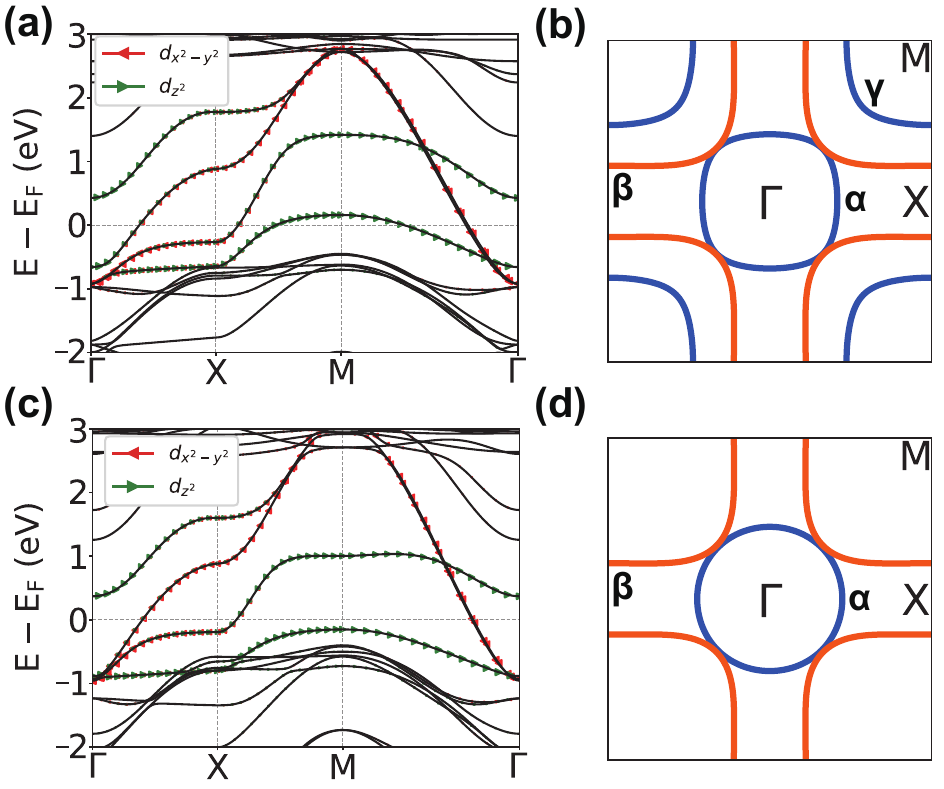}
		\caption{(a) The band structure calculated using the GGA-PBE functional at $a=b=3.83\AA$
 in the high-pressure (HP) phase. (b) The Fermi surface corresponds to the band structure in panel (a). (c) The band structure calculated using the GGA-PBE functional at $a=b=3.71\AA$
 in the thin film phase \cite{hwang}. (d) The Fermi surface corresponds to the band structure in panel (c).
			\label{figS1}}
	\end{center}
\end{figure}

Recently, superconducting \LNO thin films at ambient pressure have been successfully synthesized \cite{hwang, chen_zhuoyu}. 
They find that the superconductivity is strongly positively correlated with the reduction of the in-plane lattice constants ($a$, $b$) \cite{hwang}. Using the standard GGA-PBE functional, we calculate the band structures for different in-plane lattice constants. To simulate the scenario of substrate growth, we fix the in-plane lattice constants while allowing relaxation along the $c$-direction.

To simplify nonessential factors and highlight the underlying physics, we assume $a=b$ and adopt the crystal structure of the HP phase. The results for $a=3.83\AA$
and $a=3.71\AA$ are presented in Fig. \ref{figS1}. 
We find that as $a$ decreases, the most significant change is that the $\gamma$ band, which is typically active at the Fermi surface in previous GGA-PBE calculations \cite{yaodx,guyh}, is gradually pushed below the Fermi level, while other features of the band structure remain qualitatively unchanged.

The reason for this phenomenon is that as $a$ and $b$ decrease, the relaxed $c$-axis elongates, causing the Ni-O octahedra to compress in the $x$-, $y$-directions and stretch in the $z$-direction. This increases the Jahn-Teller distortion, which lowers the on-site energy of the \dz orbital. Although the bilayer coupling also weakens, the calculations indicate that the Jahn-Teller distortion plays the dominant role.

Therefore, even at the conventional GGA-PBE level, the calculations suggest that the emergence and enhancement of superconductivity can, to some extent, be attributed to the disappearance of the $\gamma$ band. Thus, it is reasonable to directly adopt the band structure calculated using the hybrid functional, in which the $\gamma$ band disappears near the Fermi level, as discussed in the main text.

\section{B. The fit of the TB model}

\begin{figure}
	\begin{center}
		\fig{3.4in}{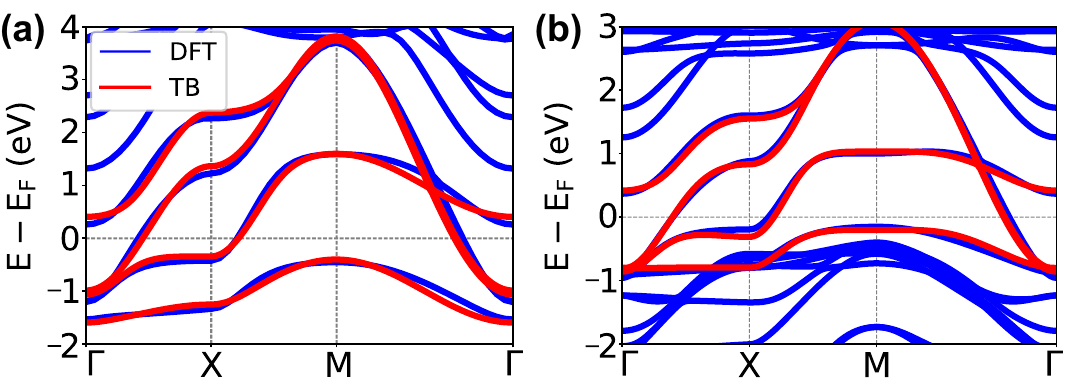}
		\caption{(a) The fit of the TB model to the HSE band structure obtained from the DFT calculations in the main text. (b) The fit of the TB model to the GGA-PBE band structure of the superconducting thin film in Fig. \ref{figS1}(c) in Appendix.
			\label{figS2}}
	\end{center}
\end{figure}

In this section, we present the fit of the TB model to the HSE band structure and GGA-PBE band structure of $a=3.71\AA$ obtained from the DFT calculations, as shown in Fig. \ref{figS2}. It can be seen that TB and DFT are in excellent agreement, indicating that our TB model provides an accurate representation of the $e_g$ orbital bands.
The TB model of $a=3.71\AA$ thin film is also listed in Table. \ref{tf_TB}. The use of the same number of parameters as in the original model in the main text cannot achieve a satisfactory fit between the TB and DFT band structures. To achieve a better fit, we incorporated additional longer-range hopping parameters into the TB model, including: intralayer same-orbital third-nearest-neighbor hopping parameter $t_{5}$, intralayer inter-orbital third-nearest-neighbor hopping parameter $t_{6}$, interlayer inter-orbital third-nearest-neighbor hopping parameter $t_{7}$ (note that the inter-orbital second-nearest-neighbor hopping parameters are zero due to symmetry constraints), and interlayer same-orbital nearest-neighbor hopping parameter $t_{\bot1}$ (with the interlayer on-site hopping parameter denoted as $t_{\bot}$). Specifically, the $T_{\textbf{k}}$, $V_{\textbf{k}}$ and $V_{\textbf{k}}$ are written as:
\begin{align}
\begin{split}
T_{\textbf{k}}^{x/z}&=t_{1}^{x/z}\gamma_{k}+t_{2}^{x/z}\alpha_{k}+t_{5}^{x/z}\delta_{k},\\
V_{\textbf{k}}&=t_{3}^{xz}\beta_{k}+t_{6}^{xz}\theta_{k},\\
V_{\textbf{k}}'&=t_{4}^{xz}\beta_{k}+t_{7}^{xz}\theta_{k},
\end{split}
\end{align}
where $\delta_{k}=2(\cos2k_{x}+\cos2k_{y})$, $\theta_{k}=2(\cos2k_{x}-\cos2k_{y})$. The matrix element originally denoted as $t_{\bot}^{x/z}$ should also be modified to $t_{\bot}^{x/z}+t_{\bot1}^{x/z}\gamma_{k}$.

\begin{table*}  
    \begin{tabular}{cccccccc}
		\hline \hline 
		
		$t_{1}^{x}$ & $t_{1}^{z}$ & $t_{2}^{x}$ & $t_{2}^{z}$ & $t_{3}^{xz}$ & $t_{4}^{xz}$ & $t_{5}^{x}$ & $t_{5}^{z}$\tabularnewline\hline 
		-0.4907 & -0.0757 & 0.0868 & -0.0123 & 0.2007 & -0.0267 & -0.0588 & -0.0148\tabularnewline \hline 
		
		$t_{6}^{xz}$ & $t_{7}^{xz}$ &  $t_{\bot}^{x}$ & $t_{\bot}^{z}$ & $t_{\bot1}^{x} $ & $t_{\bot1}^{z}$ & $\epsilon^{x}$ & $\epsilon^{z}$  \tabularnewline \hline 
		0.0274 & 0.0007 & 0.0107 & -0.6134 & 0.0039 & 0.0021 & 1.0112 & 0.2197 \tabularnewline
		
		\hline \hline
	\end{tabular}
    
    \vspace{0.5cm}
    
    \begin{tabular}{ccccccccccc}
		\hline \hline 
	sector	 &$t_{1}^{x}$ & $t_{1}^{z}$ & $t_{2}^{x}$ & $t_{2}^{z}$ & $t_{3}^{xz}$ & $t_{5}^{x}$ & $t_{5}^{z}$ & $t_{6}^{xz}$ & $\epsilon^{x}$ & $\epsilon^{z}$   \tabularnewline\hline 
	$H_{S}$&	-0.4868 & -0.0736 & 0.0868 & -0.0123 & 0.174  & -0.0588 & -0.0148 & 0.0281 & 1.0219  & -0.3937 \tabularnewline \hline
	$H_{A}$&	-0.4946 & -0.0778 & 0.0868 & -0.0123 & 0.2274  & -0.0588 & -0.0148 & 0.0267 & 1.0012 &  0.8331  \tabularnewline 
		\hline \hline
	\end{tabular}
    
	\caption{The hopping parameters of the TB Hamiltonian $H(\textbf{k})$ corresponding to the GGA-PBE results of the $a=3.71\AA$ thin film phase
    in unit of eV.  $\epsilon^{x},\epsilon^z$ are site energies for Ni $d_{x}$ and $d_{z}$ orbitals, respectively.  
	}
 \label{tf_TB}
\end{table*}

\end{document}